\documentclass[10pt,twocolumn,letterpaper]{article}

\usepackage{iccv}
\usepackage{times}
\usepackage{epsfig}
\usepackage{graphicx}
\usepackage{amsmath}
\usepackage{amssymb}

\usepackage{subcaption}
\usepackage{multirow}

\usepackage[pagebackref=true,breaklinks=true,letterpaper=true,colorlinks,bookmarks=false]{hyperref}

\iccvfinalcopy 

\def\iccvPaperID{1716} 

\ificcvfinal\fi
\begin{document}

\title{Point Cloud Super Resolution with Adversarial Residual Graph Networks}

\author{Huikai Wu~~~~Junge Zhang~~~~Kaiqi Huang\\
Institute of Automation, Chinese Academy of Sciences\\
University of Chinese Academy of Sciences\\
{\tt\small\{huikai.wu, jgzhang, kaiqi.huang\}@nlpr.ia.ac.cn}
}

\maketitle

\begin{abstract}
Point cloud super-resolution is a fundamental problem for 3D reconstruction and 3D data understanding.
It takes a low-resolution (LR) point cloud as input and generates a high-resolution (HR) point cloud with rich details.
In this paper, we present a data-driven method for point cloud super-resolution based on graph networks and adversarial losses.
The key idea of the proposed network is to exploit the local similarity of point cloud and the analogy between LR input and HR output.
For the former, we design a deep network with graph convolution.
For the latter, we propose to add residual connections into graph convolution and introduce a skip connection between input and output.
The proposed network is trained with a novel loss function, which combines Chamfer Distance (CD) and graph adversarial loss.
Such a loss function captures the characteristics of HR point cloud automatically without manual design.
We conduct a series of experiments to evaluate our method and validate the superiority over other methods.
Results show that the proposed method achieves the state-of-the-art performance and have a good generalization ability to unseen data.
Code is available at \url{https://github.com/wuhuikai/PointCloudSuperResolution}.
\end{abstract}

\section{Introduction}
\begin{figure}
\begin{center}
	\subcaptionbox{Input}{\includegraphics[width=0.24\linewidth]{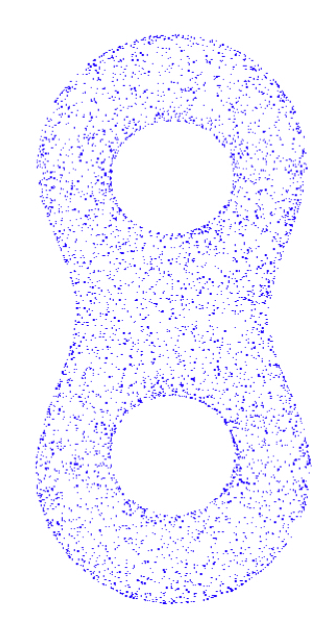}}
	\subcaptionbox{GT}{\includegraphics[width=0.24\linewidth]{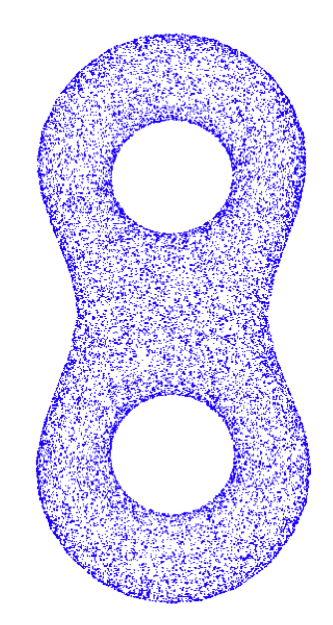}}
	\subcaptionbox{PU-Net~\cite{yu2018pu}}{\includegraphics[width=0.24\linewidth]{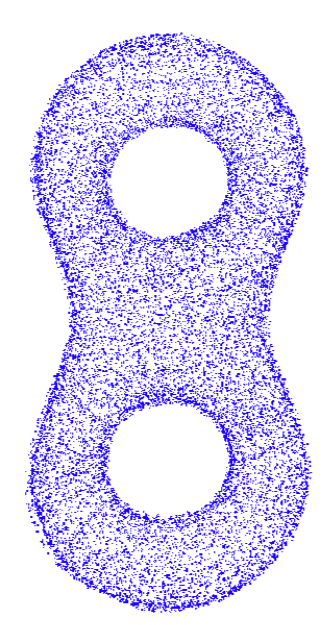}}
	\subcaptionbox{Ours}{\includegraphics[width=0.24\linewidth]{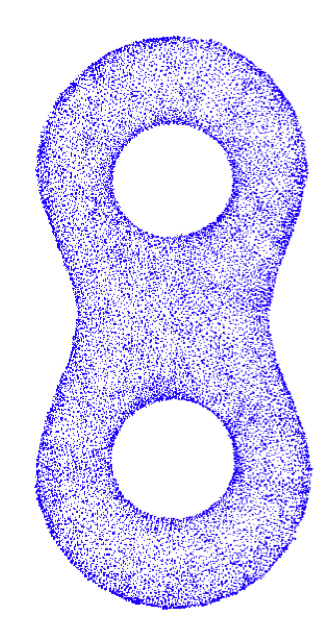}}\\
\end{center}
	\vspace{-1.5em}
	\caption{\textbf{Point Cloud Super Resolution.} (a) is the input LR point cloud with sparse distribution. (b) is the corresponding HR point cloud with dense distribution. (c) and (d) are the HR point cloud generated by PU-Net~\cite{yu2018pu} and our method respectively. Ours is sharper at edges with fewer noisy points. Best viewed in color.}
	\vspace{-1em}
	\label{fig:pc_sr}
\end{figure}

When modeling an object from the real world for 3D printing or animation, a common way is to first obtain the point cloud with depth scanning devices or 3D reconstruction algorithms~\cite{fan2017point,hartley2003multiple} and then recover the mesh from the point cloud~\cite{bernardini1999ball}.
However, the captured point cloud is usually sparse and noisy due to the restrictions of devices or the limitations of algorithms, which leads to a low-quality mesh.

The key to improving the quality of the recovered mesh is point cloud super-resolution, which takes a LR point cloud as input and generates a HR point cloud with rich details and few noisy points, as shown in Figure~\ref{fig:pc_sr}.
Most existing methods are optimization based without learning from data, which have strong assumptions about the underlying surface of the HR point cloud~\cite{huang2009consolidation,huang2013edge}.
As for data-driven methods, few prior works study deep learning on this problem.
\cite{yu2018pu} propose PU-Net for point cloud super-resolution, which is a pioneering work.
By employing deep neural networks, it outperforms many traditional methods such as EAR~\cite{huang2013edge} and achieves the state-of-the-art performance.

In this paper, we aim to advance the performance of point cloud super-resolution by overcoming the defects of PU-Net.
The first problem is that PU-Net directly regresses the point coordinates without exploiting the similarity between LR and HR point cloud, which makes it hard to train.
The second problem is that PU-Net proposes a complicated loss function with a strong assumption on the uniform distribution of HR point cloud.
Manually designed loss functions tend to overfit human priors, which fail to capture many other properties of HR point cloud, such as continuity.

Recent work~\cite{kim2016deeply} in image super-resolution shows that predicting the residual between the LR and HR image is a more desirable way to achieve better accuracy.
Thus, to solve the first problem, we propose to introduce residual connections into graph convolution networks (GCNs)~\cite{defferrard2016convolutional} and add a skip connection between the input layer and the output layer.
Employing GCNs to process point cloud is not new~\cite{zhang2018graph}.
However, the GCN in our method is unique in two aspects compared to that in \cite{zhang2018graph}: (1) The architecture of our GCN is designed for generating point cloud while that in \cite{zhang2018graph} aims at aggregating information for classification. (2) We propose an un-pooling layer for the GCN to upsample the input point cloud.

To solve the second problem, we design a graph adversarial loss based on LS-GAN~\cite{mao2017least}.
The proposed loss function is more expressive than manually designed ones, which can capture the characteristics of HR point cloud automatically.
Pan~\etal also introduce adversarial loss into graph networks~\cite{Pan2018AdversariallyRG}.
However, they focus on learning the distribution of graph embeddings.
Thus, a multi-layer perceptron is employed as the discriminator to process the input vector.
Differently, we aim at distinguishing real and fake point clouds.
To achieve this, we propose a GCN as the discriminator to process the generated point cloud, which is significantly different from \cite{Pan2018AdversariallyRG}.

In this way, we propose a novel method for point cloud super-resolution, named Adversarial Residual Graph Convolution Network (AR-GCN).
Experiments show that the proposed method achieves the state-of-the-art performance on both seen dataset~\cite{yu2018pu} and unseen dataset (SHREC15).
The contributions of our method are three-folds.
First, we propose a novel architecture for point cloud super-resolution.
Second, we introduce the graph adversarial loss to replace manually designed loss functions.
Third, we advance the state-of-the-art performance on both seen and unseen datasets.

\section{Related Work}
\subsection{Point Cloud Super Resolution}
Point cloud super-resolution is formulated as an optimization problem in most earlier works.
To upsample a point cloud, \cite{alexa2003computing} first compute the Voronoi diagram on the moving least squares surface and then add points at the vertices of this diagram.
\cite{lipman2007parameterization} present a locally optimal projection operator for surface approximation based on $L_1$ median, which is parameter-free and robust to noisy points.
These methods have a strong assumption on the smoothness of the underlying surface, which tend to have vague edges.
Thus, \cite{huang2013edge} introduce an edge-aware point cloud upsampling method, which first samples away from the edges and then progressively samples the point cloud while approaching the edge singularities.

However, all these methods have strong assumptions about the underlying surface based on human priors.
To exploit the massive 3D data, \cite{yu2018pu} propose a data-driven method that first learns multi-level features per point with PointNet++~\cite{qi2017pointnet++} and then expands the point set via a multi-branch convolution.
Through end-to-end learning, this method outperforms the optimization based methods on multiple datasets, which achieves the state-of-the-art performance.

\subsection{Deep Learning for 3D Data}
There is a rising interest in 3D data processing recently.
Most existing works transform 3D data into volumetric grids, which are then processed by 3D CNNs~\cite{smith20183d,maturana2015voxnet,wu20153d,riegler2017octnet}.
Particularly, \cite{smith20183d} propose to upsample 3D objects in voxel space, which is similar to point cloud super-resolution.

3D CNNs are memory and time consuming.
Thus, following works propose to process point cloud directly instead of the volumetric grid~\cite{qi2016pointnet,qi2017pointnet++,li2018pointcnn,su2018splatnet,li2018so,jiang2018pointsift}.
\cite{qi2016pointnet} propose a point-wise network for 3D object classification and segmentation.
Successively, \cite{qi2017pointnet++} introduce a hierarchical feature learning architecture to capture local and global context.
\cite{li2018pointcnn} introduce a convolution operator named X-conv that is capable of leveraging spatially-local correlation for point cloud  classification.

Besides 3D scene understanding, multiple works focus on employing deep learning for single image 3D reconstruction~\cite{fan2017point,wang2018pixel2mesh,sun2018pix3d,kato2018neural}.
\cite{fan2017point} present a conditional shape sampler to generate multiple plausible point clouds from a single image.
Differently, \cite{wang2018pixel2mesh} propose a graph CNN for producing the 3D triangular mesh from a single color image.

\subsection{Image Super Resolution}
Super-resolution for 2D images is a well-studied problem, which is closely related to point cloud super-resolution.
Modern approaches are usually built on deep learning in a data-driven manner~\cite{dong2014learning, wang2016end, ledig2017photo, zhang2018residual, han2018image, haris2018deep}.
\cite{dong2014learning} first upsample the low-resolution image with bicubic interpolation and then use a shallow CNN to recover the details and textures.
Instead of manually designed interpolation, \cite{wang2016end} propose to jointly learn the interpolation and detail recovery.
To generate realistic images, \cite{ledig2017photo} present a generative adversarial network, which pushes the generated images close to natural images.

\section{Method}
\begin{figure*}
\begin{center}
    \includegraphics[width=0.95\linewidth]{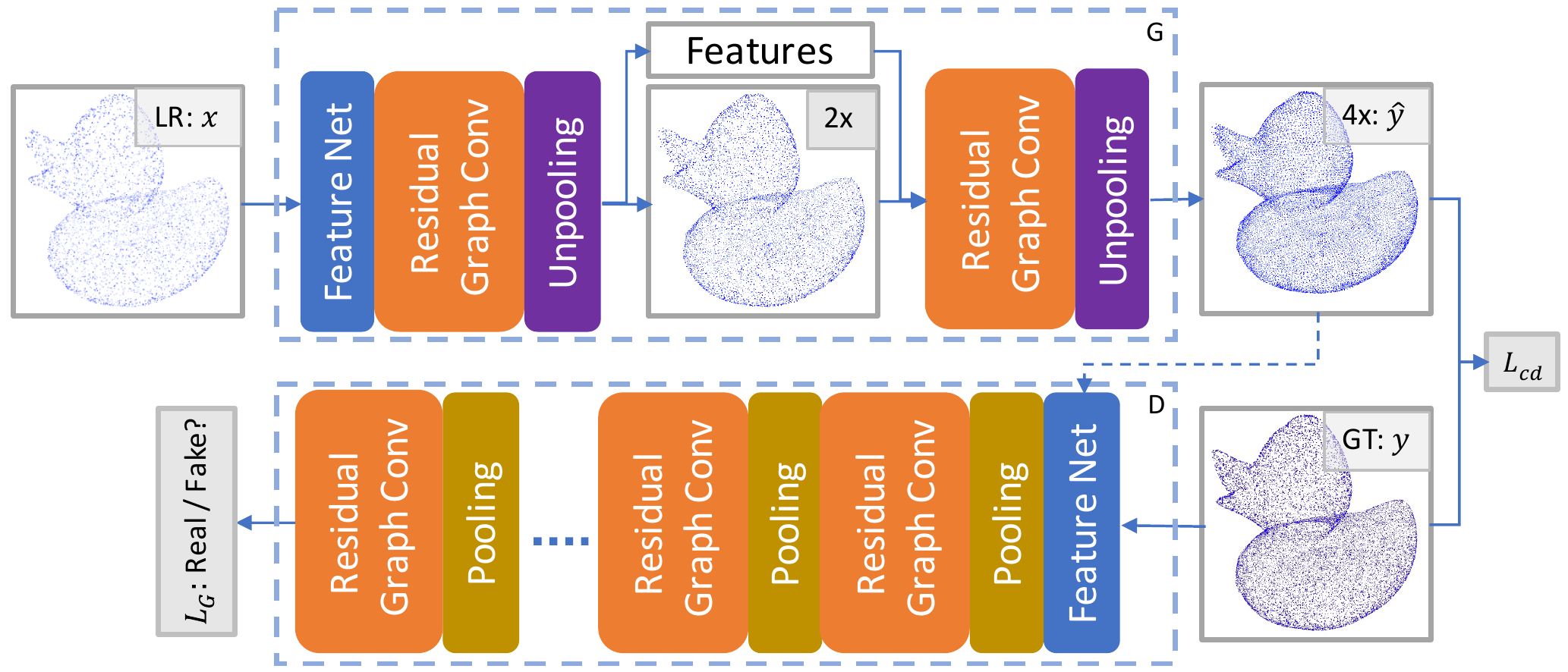}
\end{center}
	\vspace{-1.5em}
	\caption{\textbf{Framework Overview.} The proposed AR-GCN consists of a generator $G$ and a discriminator $D$. $G$ is a residual GCN that upsamples the input point cloud progressively with the upsampling ratio $2\times$. $D$ is also a residual GCN, which learns to distinguish fake HR point cloud from the real one.}
	\vspace{-1em}
	\label{fig:framework}
\end{figure*}

In this section, we first define point cloud super-resolution formally and then introduce our method AR-GCN in detail.
Our method is a novel approach that contains three major components: the adaptive adversarial loss function $L_G$, the residual GCN $G$, and the graph discriminator $D$.
As shown in Figure~\ref{fig:framework}, the LR point cloud $x$ is directly fed into $G$ to generate the HR output $\hat{y}$.
Then, $\hat{y}$ is sent into $D$ to produce $L_G$, while another loss $L_{cd}$ is calculated based on $\hat{y}$ and the ground truth $y$.

\subsection{Point Cloud Super Resolution}
Formally, given a point cloud $x$ with shape $n\times3$, the goal of point cloud super-resolution is to generate a point cloud $\hat{y}$ with shape $N\times3$ ($N = \gamma n, \gamma > 1$).
Each point of $\hat{y}$ lies on the underlying surface described by $x$, as shown in Figure~\ref{fig:pc_sr}.

\subsection{Method Overview: AR-GCN}
As shown in Figure~\ref{fig:framework}, our method AR-GCN consists of two networks, the generator $G$ and the discriminator $D$.
$G$ aims to generate the HR point cloud by upsampling the LR input progressively, while $D$ is responsible for distinguishing the fake HR point cloud from the real one.
To train $G$ and $D$ simultaneously, we propose a joint loss function as shown in Equation~\ref{eq:loss}:
\begin{equation}
    L(x, y) = \lambda L_{cd}(G(x), y) + L_G(G(x)),
    \label{eq:loss}
\end{equation}
where $\lambda$ controls the balance between $L_{cd}$ and $L_{G}$.

$L_{cd}$ measures the distance between $y$ and $\hat{y}$, which is similar to $L_2$ loss in image super-resolution.
As shown in Equation~\ref{eq:cd}:
\begin{equation}
    L_{cd}(\hat{y}, y) = \Sigma_{p \in y} min_{q \in \hat{y}}{||p-q||^2_2},
    \label{eq:cd}
\end{equation}
which is a variant of Chamfer Distance.
The original Chamfer Distance consists of two parts: $L_{cd}$ and $L_{\hat{cd}}$, where $L_{\hat{cd}}$ is symmetric with $L_{cd}$ and defined as Equation~\ref{eq:cd_1}:
\begin{equation}
    L_{\hat{cd}}(\hat{y}, y) = \Sigma_{q \in \hat{y}} min_{p \in y}{||p-q||^2_2}.
    \label{eq:cd_1}
\end{equation}
$L_{\hat{cd}}$ encourages $\hat{y}$ to be identical to the LR input, which leads to duplication points in the output point cloud.
Thus, we remove $L_{\hat{cd}}$ and only employ $L_{cd}$ as our loss function.

$L_{cd}$ measures the point-wise distance between $y$ and $\hat{y}$, which ignores high-order properties defined by a cluster of points, such as continuity.
Traditional methods usually manually design a complex function as the loss, which is inefficient and has strong assumptions about the underlying surface.
Alternatively, we propose a loss function $L_G$ that is defined by a network and learned from data automatically.
Concretely, $L_G$ is a graph adversarial loss that is inspired by generative adversarial networks (GANs)~\cite{goodfellow2014generative}.
In this paper, we employ LS-GAN~\cite{mao2017least} as the adversarial loss for its simplicity and effectiveness.
$L_G$ is defined as follows:
\begin{equation}
    L_G(\hat{y}) = ||1 - D(\hat{y})||_2^2, 
\end{equation}
where $D$ is the discriminator that aims to distinguish real and fake HR point cloud by minimizing the following loss:
\begin{equation}
    L_D(\hat{y}, y) = \frac{1}{2}||D(\hat{y})||_2^2 + \frac{1}{2}||1 - D(y)||_2^2.
\end{equation}

\subsection{Residual Graph Convolution Generator}
The generator $G$ is built on the Graph Convolution Network (GCN)~\cite{bronstein2017geometric}, which aims to progressively upsample the LR point cloud.
It consists of three building blocks, namely residual graph convolution block, unpooling block and feature net, as shown in Figure~\ref{fig:framework}.

\subsubsection{Residual Graph Convolution Block}
\begin{figure}
\begin{center}
    \includegraphics[width=0.9\linewidth]{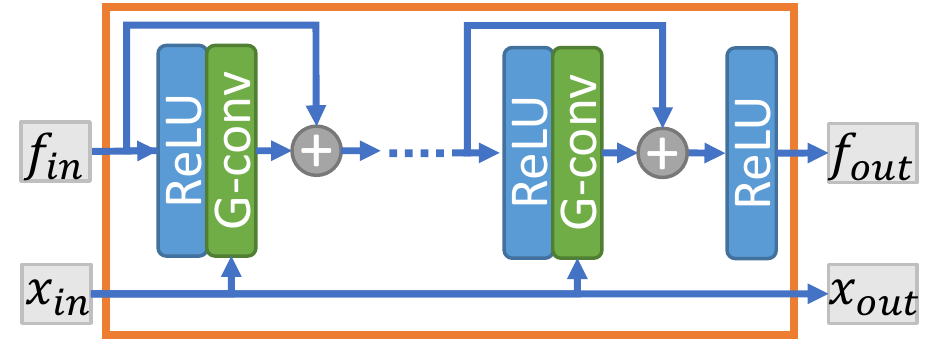}
\end{center}
	\vspace{-1.5em}
	\caption{\textbf{Residual Graph Convolution Block.} $x_{in}$ is used for querying the $k$ nearest neighbors. $x_{out}$ is the same as $x_{in}$.}
	\vspace{-1em}
	\label{fig:rgcn}
\end{figure}

~~~PU-Net employs PointNet++ to generate HR point cloud, which treats the central point and the neighbor points equally.
This limits the learning ability of the network.
Alternatively, we build our method on graph convolution~\cite{bronstein2017geometric}, as shown in Figure~\ref{fig:rgcn}.

The core of graph convolution, G-conv, is defined on a graph G=($\upsilon$, $\varepsilon$) and calculated as follows,
\begin{equation}
    f^p_{l+1} = w_0f^p_l + w_1\Sigma_{q\in N(p)}f^q_l, \forall p \in \upsilon,
\end{equation}
where $w$ is the learn-able parameters and $f^p_l$ represents the feature of vertex $p$ at layer $l$.
$N(p)$ is the vertices that connect to $p$ as defined by the adjacency matrix $\varepsilon$.
However, there's no predefined adjacency matrix for a point cloud.
To solve this problem, we define $N(p)$ as the $k$ nearest neighbors of $p$ in Euclidean space, of which the coordinates are defined by $x_{in}$.

Besides G-conv, we also introduce residual connections into our block, because residual networks usually lead to faster convergence and better results.
It also helps to exploit the similarity between LR point cloud and the corresponding HR point cloud.

In our experiment, the number of neighbors $k$ is set to 8.
All the G-conv operators inside the block have the same number of channels, which is 128.
The input feature $f_{in}$ and point cloud $x_{in}$ are processed by 12 residual layers to obtain $f_{out}$, while $x_{out}$ is the same as $x_{in}$.

\subsubsection{Unpooling Block}
~~~~The unpooling block takes point cloud $x_{in}$ and the corresponding features $f_{in}$ as inputs.
It first transforms $f_{in}$ with shape $\hat{n}\times c$ to a tensor with shape $\hat{n} \times 6$ by a G-conv layer.
The tensor is then reshaped to $\hat{n}\times 2\times 3$, which is noted as $\delta x$.
The upsampled point cloud $x_{out}$ is obtained by adding $x_{in}$ and $\delta x$ point-wisely, where each point is transformed into 2 points.
The unpooling block is designed to predict the residual between $x_{in}$ and $x_{out}$ instead of regressing $x_{out}$ directly.
This exploits the similarity between $x_{in}$ and $x_{out}$, which leads to a faster convergence and better performance.

The feature of the output point cloud, $f_{out}$, are obtained by the following equation:
\begin{equation}
    f_{out}^p = \frac{1}{k}\Sigma_{q\in N[x_{in}](p)}f_{in}^q, \forall p \in x_{out},
\end{equation}
where $N[x_{in}](p)$ means the $k$ nearest neighbors of point $p$ in point cloud $x_{in}$.

\subsubsection{Feature Net}
~~~~As shown in Figure~\ref{fig:rgcn}, a residual graph convolution block takes both the point cloud and the corresponding feature as inputs.
However, the generator only have one input, the point cloud $x$.
To obtain the other input, the corresponding feature $f$, we design a simple block named feature net, which takes the point cloud $x$ as input.
Specifically, for each point $p\in x$ with shape $1\times 3$, we first obtain its $k$ nearest neighbors $P$ with shape $k\times 3$.
Then, a series of point-wise convolutions with a max-pooling layer transform $\hat{P} = P - p$ into $f^p$ with shape $1\times c$.

In our experiment, $k$ is set to 8 while $c$ is set to 128.
The number of convolution layers is set to 3.

\subsubsection{Progressive Super Resolution}
~~~~~Instead of directly upsampling the LR point cloud with the desired upscale ratio, we choose to generate the HR point cloud step by step.
The point cloud is upsampled by 2 times at each step, as shown in Figure~\ref{fig:framework}.
Our experiment shows that such an approach results in better accuracy.

\begin{table*}[ht]
\small
\begin{center}
\begin{tabular}{l|cc|c|ccccc|cc|c}
\hline
\multirow{2}{*}{Method} & \multirow{2}{*}{CD} & \multirow{2}{*}{EMD} & \multicolumn{1}{c}{F-score} & \multicolumn{5}{|c}{NUC with different p} & \multicolumn{2}{|c|}{Deviation (1e-2)} & \multirow{2}{*}{Params}\\
\cline{4-11}
& & & $\tau=0.01$ & $0.2\%$ & $0.4\%$ & $0.6\%$ & $0.8\%$ & $1.0\%$ & mean & std \\
\hline
\hline
Input & 0.0120 & 0.0036 & 41.33\%  & 0.315 & 0.224 & 0.185 & 0.163 & 0.150 & - & - & -\\
MLS~\cite{alexa2003computing} & 0.0117 & 0.0043 & 57.70\% & 0.364 & 0.272 & 0.229 & 0.204 & 0.186 & \textbf{0.18} & 0.34 & -\\
PU-Net~\cite{yu2018pu} & 0.0118 & 0.0041 & 43.24\% & 0.206 & 0.165 & 0.147 & 0.137 & 0.131 & 0.78 & 0.66 & \textbf{0.777M}\\
\hline
AR-GCN & \textbf{0.0084} & \textbf{0.0035} & \textbf{70.28}\% & \textbf{0.204} & \textbf{0.164} & \textbf{0.145} & \textbf{0.134} & \textbf{0.128} & 0.26 & \textbf{0.30} & 0.785M\\
\hline
\end{tabular}
\end{center}
	\vspace{-1.5em}
	\caption{\textbf{Quantitative Comparison on the Train-Test Dataset.}}
	\label{table:pu_result}
\end{table*}

\begin{table*}[ht]
\begin{center}
\begin{tabular}{l|cc|c|ccccc|cc}
\hline
\multirow{2}{*}{Method} & \multirow{2}{*}{CD} & \multirow{2}{*}{EMD} & \multicolumn{1}{c}{F-score} & \multicolumn{5}{|c}{NUC with different p} & \multicolumn{2}{|c}{Deviation (1e-2)}\\
\cline{4-11}
& & & $\tau=0.01$ & $0.2\%$ & $0.4\%$ & $0.6\%$ & $0.8\%$ & $1.0\%$ & mean & std \\
\hline
\hline
Input & 0.0077 & \textbf{0.0031} & 27.86\% & 0.310 & 0.220 & 0.183 & 0.163 & 0.151 & - & - \\
MLS~\cite{alexa2003computing} & 0.0067 & 0.0032 & 84.69\% & 0.253 & 0.199 & 0.173 & 0.159 & 0.150 & 0.33 & 0.46\\
PU-Net~\cite{yu2018pu} & 0.0103 & 0.0050 & 56.39\% & 0.283 & 0.230 & 0.204 & 0.189 & 0.180 & 0.90 & 0.73\\
\hline
AR-GCN & \textbf{0.0054} & \textbf{0.0031} & \textbf{93.07}\% & \textbf{0.201} & \textbf{0.162} & \textbf{0.144} & \textbf{0.135} & \textbf{0.130} & \textbf{0.18} & \textbf{0.19}\\
\hline
\end{tabular}
\end{center}
	\vspace{-1.5em}
	\caption{\textbf{Quantitative Comparison on SHREC15.}}
	\vspace{-1em}
	\label{table:shrec_result}
\end{table*}

\subsection{Graph Discriminator}
To generate more realistic HR point cloud, we present a graph adversarial loss for point cloud, which is defined by the discriminator $D$.
As shown in Figure~\ref{fig:framework}, $D$ is composed of feature net, residual graph convolution block, and pooling block.
For feature net, $k$ is set to 8 while $c$ is set to 64.
The number of convolution layers is set to 2.
For residual graph convolution block, $k$ is set to 8 while $c$ is set to 64.
The number of layers is set to 4.

\vspace{-1em}
\paragraph{Pooling Block}
Given the input point cloud $x_{in}$ with shape $4n\times3$, we first employ farthest point sampling (FPS) to generate $x_{out}$ with shape $n\times3$.
The corresponding features $f_{out}$ is then obtained as follows:
\begin{equation}
    f_{out}^p = max_{q\in N[x_{in}](p)}f_{in}^q, \forall p \in x_{out},
\end{equation}
where $max$ is applied element-wisely.

\vspace{-1em}
\paragraph{Graph Patch GAN}
Most discriminators downsample the input progressively to obtain a single flag for the whole input.
Such a design usually leads to blurry and unpleasant artifacts.
Instead of employing a global discriminator, we build a graph patch GAN based on~\cite{shrivastava2017learning}.
Specifically, our discriminator downsamples the input multiple times so that the output contains more than 1 point.
Graph patch GAN forces every local patch of the generated point cloud to lie on the distribution of the real HR point cloud.
In our experiment, we set the number of the output points to 64.

\section{Experiment}
In this section, we first introduce the datasets for training and testing, as well as the details of our implementation.
The quantitative and qualitative results are then presented to show the effectiveness of our method.
To demonstrate the effect of each component in AR-GCN, we also conduct a comprehensive ablation study.
To further show the potential applications of our method, we test AR-GCN in an iterative setting and apply it to 3D reconstruction task.
Besides, we also employ our method to assist LR point cloud classification.

\subsection{Datasets}
We utilize two datasets for our experiments.
One is the train-test dataset, which our method is trained with and tested on.
The other is the unseen dataset, where our method is directly tested without training or finetuning.

\vspace{-1em}
\paragraph{Train-Test Dataset}
We use the dataset proposed in PU-Net~\cite{yu2018pu} for training and testing.
This dataset contains $60$ different models from the Visionair repository.
Following the protocol in PU-Net, we use $40$ models for training while the other $20$ models are used for testing.
For training, $100$ patches are extracted from each model as the ground truth, which contains $4,096$ points.
The input patch is randomly sampled from the ground truth patch at each iteration of training, which contains $1,024$ points.
For testing, we sample $20,000$ points uniformly per model as the ground truth, while sampling $5,000$ points as the input.

\vspace{-1em}
\paragraph{Unseen Dataset: SHREC15} To further validate the generalization ability of our method, we directly test AR-GCN on SHREC15~\cite{Lian:2015:NSR:2852282.2852307} after training with the train-test dataset without finetuning.
For SHREC15, there are 50 categories in total and 24 models in each category.
We randomly choose one model from each category for testing, since the models in each category only differ in the pose.
Same as the train-test dataset, the ground truth contains $20,000$ points, while the input contains $5,000$ points.

\subsection{Implementation Details}
\label{sec:training}
Our method is implemented in Tensorflow~\cite{abadi2016tensorflow} and runs on a single Titan-Xp GPU.
To avoid overfitting, we augment the training data by randomly rotating, shifting and scaling the data.
For optimization, we use Adam~\cite{kingma2014adam} as the optimizer, where the batch size is 28 and the learning rate is 0.001.
The network is firstly trained with $L_{cd}$ for 80 epochs at the speed of $2.2$ min/epoch.
Then we add $L_G$ and finetune the network for another 40 epochs at the speed of $4.9$ min/epoch.
The training process takes about $6.2$ hours in total, while it takes about $4.8$ hours for training PU-Net under the same setting ($2.4$ min/epoch).

\begin{figure*}
\captionsetup[subfigure]{labelformat=empty}
\begin{center}
	\subcaptionbox{}{\includegraphics[width=0.2\linewidth]{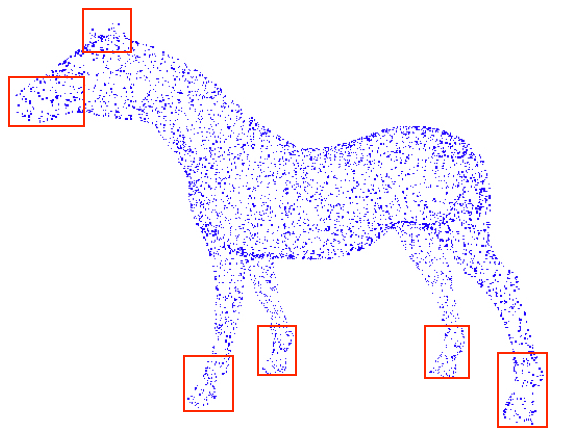}}
	\hfill
	\subcaptionbox{}{\includegraphics[width=0.2\linewidth]{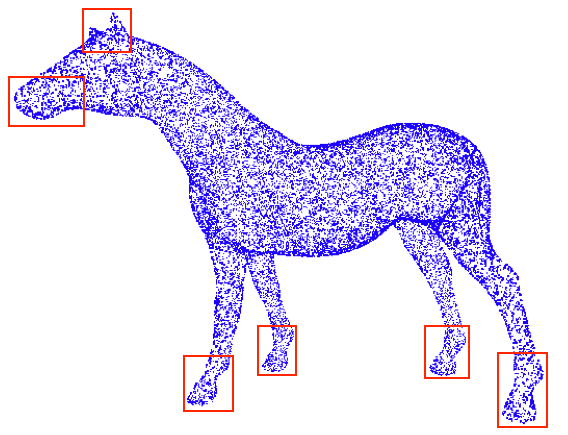}}
	\hfill
	\subcaptionbox{}{\includegraphics[width=0.2\linewidth]{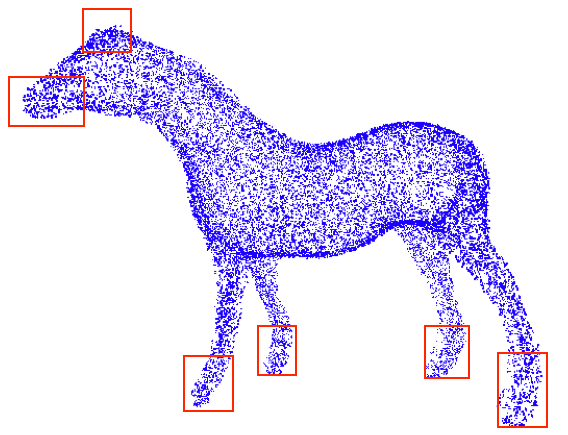}}
	\hfill
	\subcaptionbox{}{\includegraphics[width=0.2\linewidth]{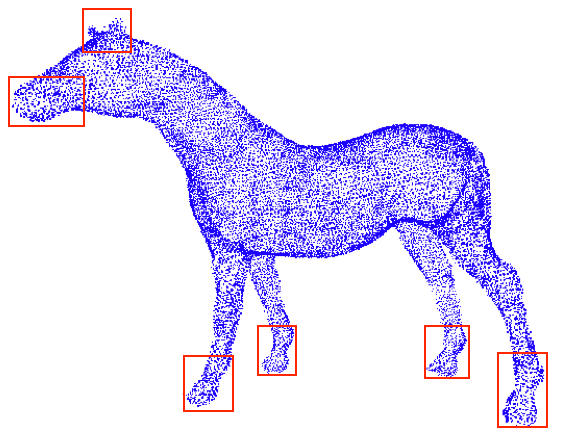}}\\
	\subcaptionbox{(a) Input}{\includegraphics[width=0.2\linewidth]{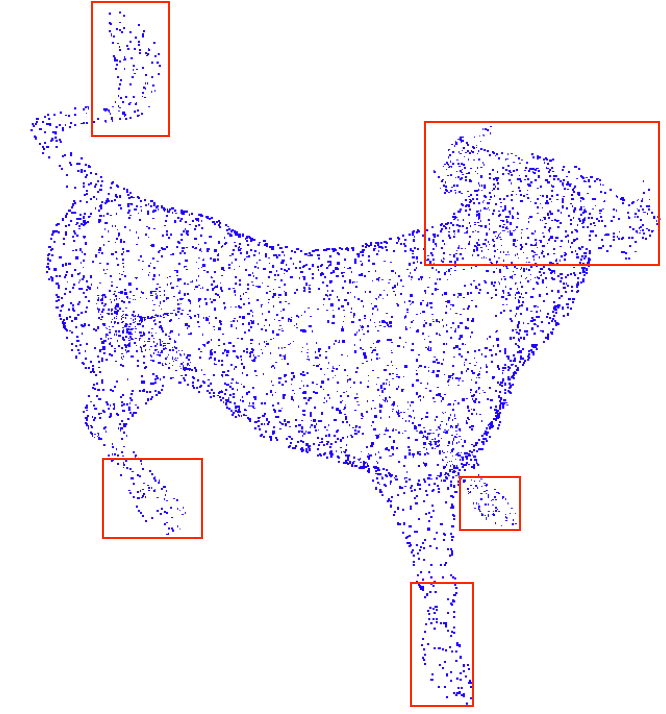}}
	\hfill
	\subcaptionbox{(b) GT}{\includegraphics[width=0.2\linewidth]{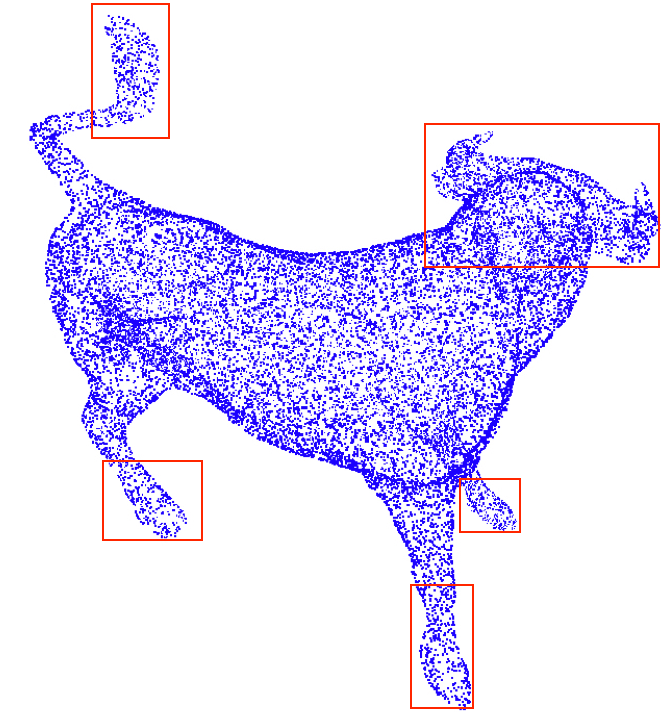}}
	\hfill
	\subcaptionbox{(c)
	PU-Net~\cite{yu2018pu}}{\includegraphics[width=0.2\linewidth]{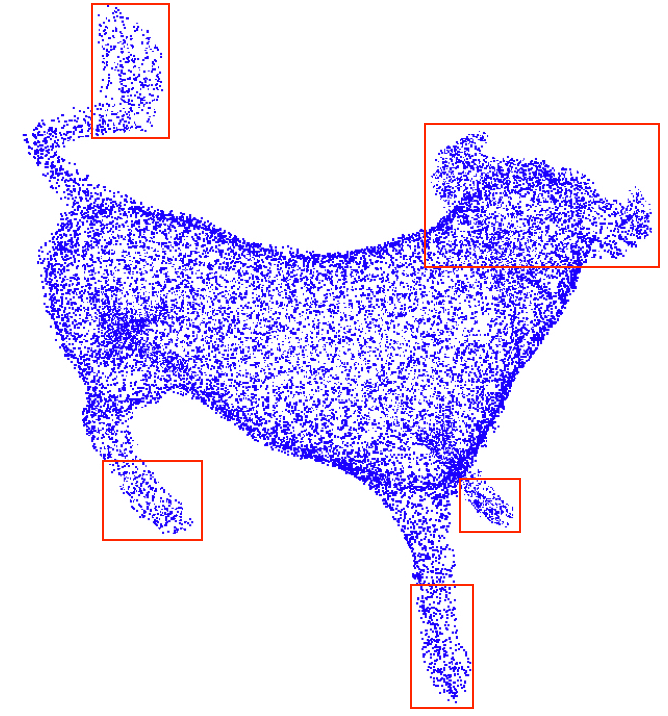}}
	\hfill
	\subcaptionbox{(d) Ours}{\includegraphics[width=0.2\linewidth]{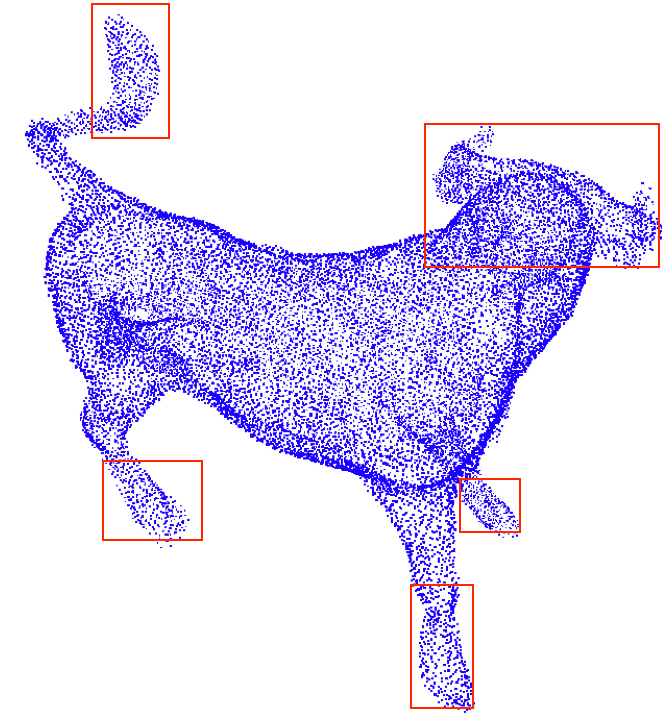}}\\
\end{center}
	\vspace{-1.5em}
	\caption{\textbf{Qualitative Results.} (a) is the LR input point cloud with a sparse distribution, while (b) is the corresponding HR point cloud with a dense distribution. (c) and (d) are the HR point cloud generated by PU-Net~\cite{yu2018pu} and our method. Our results are sharper at edges, which have richer details but fewer noisy points, especially in the red boxes. Best viewed in color.}
	\label{fig:visual}
\end{figure*}

\subsection{Quantitative Results}
\paragraph{Evaluation Metrics}
We adopt standard 3D reconstruction metrics for point cloud super-resolution because the output of both tasks is point cloud.
To measure the difference between $\hat{y}$ and $y$ point-wisely, we utilize the standard Chamfer Distance (CD) and Earth Mover's Distance (EMD), for which smaller is better.

CD and EMD are heavily influenced by the outliers.
Thus, we also report F-score~\cite{sokolova2006beyond} by treating point cloud super-resolution as a classification problem.
Specifically, precision and recall are first evaluated by checking the percentage of points in $\hat{y}$ or $y$ that can find a neighbor from the other within certain threshold $\tau$.
The F-score is then calculated as the harmonic mean of precision and recall.
For this metric, larger is better.

The metrics in \cite{yu2018pu} are also employed to evaluate our method.
We use Deviation to measure the difference between the predicted point cloud and the ground truth mesh, while normalized uniformity coefficient (NUC) is evaluated for measuring the uniformity.
For these two metrics, smaller is better.
Notably, the original mesh is used as the ground truth instead of the sampled $20,000$ points.

\begin{table*}[ht]
\small
\begin{center}
\begin{tabular}{l|cc|c|cc||cc|c|cc}
\hline
\multirow{3}{*}{Method} & \multicolumn{5}{c||}{Train-Test Dataset} & \multicolumn{5}{c}{SHREC15} \\
\cline{2-11}
& \multirow{2}{*}{CD} & \multirow{2}{*}{EMD} & \multicolumn{1}{|c}{F-score} & \multicolumn{2}{|c||}{Deviation (1e-2)} & \multirow{2}{*}{CD} & \multirow{2}{*}{EMD} & \multicolumn{1}{|c}{F-score} & \multicolumn{2}{|c}{Deviation (1e-2)} \\
\cline{4-6}
\cline{9-11}
& & & $\tau=0.01$ & mean & std & & & $\tau=0.01$ & mean & std \\
\hline
\hline
$\textnormal{GCN}_{points}^{4\times}$ & 0.0971 & 0.0857 & ~3.96\% & 13.7 & 12.3 & 0.0925 & 0.0778 & ~8.03\% & 10.3 & 10.1\\
$\textnormal{GCN}^{4\times}$ & 0.0092 & 0.0036 & 63.79\% & 0.31 & 0.35 & 0.0061 & 0.0031 & 87.98\% & 0.24 & 0.27 \\
$\textnormal{ResGCN}^{4\times}$ & 0.0090 & 0.0036 & 65.04\% & 0.36 & 0.37 & 0.0059 & 0.0031 & 90.57\% & 0.25 & 0.25 \\
\hline
PU-Net~\cite{yu2018pu} & 0.0118 & 0.0041 & 43.24\% & 0.78 & 0.66 & 0.0103 & 0.0050 & 56.39\% & 0.90 & 0.73\\
ResGCN & 0.0086 & \textbf{0.0035} & 68.75\% & 0.30 & 0.31 & 0.0056 & 0.0031 & 92.32\% & 0.21 & 0.21\\
ResGCN + $\textnormal{L}_{pu}$ & 0.0096 & 0.0040 & 60.71\% & 0.36 & 0.50 & 0.0064 & 0.0031 & 86.46\% & 0.26 & \textbf{0.14} \\
\hline
AR-GCN w/o FT & 0.0085 & 0.0041 & 69.53\% & 0.27 & \textbf{0.30} & 0.0055 & \textbf{0.0027} & 92.46\% & \textbf{0.18} & 0.19 \\
AR-GCN & \textbf{0.0084} & \textbf{0.0035} & \textbf{70.28}\% & \textbf{0.26} & \textbf{0.30} & \textbf{0.0054} & 0.0030 & \textbf{93.07}\% & \textbf{0.18} & 0.19 \\
\hline
\end{tabular}
\end{center}
	\vspace{-1.5em}
	\caption{\textbf{Ablation Study Results on the Train-Test Dataset and the Unseen Dataset, SHREC15.}}
	\vspace{-1em}
	\label{table:ablation_result}
\end{table*}

\vspace{-1em}
\paragraph{Comparison with Other Methods}
The performance of different methods on the train-test dataset is reported in Table~\ref{table:pu_result}.
We first report the performance of the LR input as a preliminary baseline.
We then report the performance of Moving Least Squares (MLS)~\cite{alexa2003computing}, which is a traditional method.\footnote{The result of MLS is reproduced with Point Cloud Library (PCL).}
The performance of PU-Net is then reported, which is the state-of-the-art method for point cloud super-resolution.\footnote{The result of PU-Net is reproduced with the official code by following the author's instructions.}

As shown in Table~\ref{table:pu_result}, our method outperforms all the other methods under most metrics.
Particularly, our method improves the F-score by more than $10\%$, while advances CD a large step.
Surprisingly, our method even outperforms PU-Net in NUC, although it is not trained with the repulsion loss~\cite{yu2018pu}, which forces the generated point cloud uniform.
We attribute this to the effect of our adversarial loss.
As for MLS, it performs well in Deviation but has the lowest NUC scores.
The reason is that MLS tends to produce new points close to the points in the input point cloud, which leads to a non-uniformly distributed point cloud, resulting in poor performance on NUC metric.
However, the mean deviation to the ground truth is small.
Similar results are obtained on SHREC15 as shown in Table~\ref{table:shrec_result}, which shows the generalization ability of our method on the unseen dataset.
As for NUC and Deviation, our method outperforms the baseline methods by a large margin.

Since both our method and PU-Net are deep learning based method, we also compare the number of parameters in each model.
As shown in Table~\ref{table:pu_result}, our model contains about the same number of learnable parameters as PU-Net while achieves much better performance.

\subsection{Qualitative Results}
The evaluation metrics reflect the shape quality to some degree.
However, they mainly focus on point-wise distance and fail to reflect the surface properties such as smoothness and high-order details.
Since there are barely any standard metrics to measure these aspects, we present a series of qualitative results to show the advantage of our method.

Figure~\ref{fig:visual} presents the point clouds in both datasets visually, where the 1st row is from the train-test dataset and the 2nd row is from SHREC15.
Compared to PU-Net, the HR point clouds of our method have richer details and sharper edges.
Besides, ours are more uniformly distributed in the smooth area with fewer noisy points.
On the contrary, the results from PU-Net are noisy and blurry on the edges with little details.
It tends to outspread uniformly without preserving the underlying structure.
Notably, the differences around the legs, feet and horns are most obvious, as shown in the red boxes.

\subsection{Ablation Study}
To further verify the effect of each component in our method, we conduct an ablation study on both datasets and present the results in Table~\ref{table:ablation_result}.

We first present the results of a very simple baseline $\textnormal{GCN}_{points}^{4\times}$.
Compared to AR-GCN, there are mainly three modifications.
First, we remove the residual connections from the proposed generator, as well as the skip connection between input and output.
Second, it regresses the coordinates directly under the supervision of a single loss $\textnormal{L}_{cd}$.
Third, instead of progressively upscaling, it upsamples the point cloud by $4\times$ directly with the number of G-conv layers unchanged.
Without all the key features in our method, it is not surprising that this simple baseline does not work at all.
We then put back the skip connection between input and output, which forces the model to predict the residual $\delta x$ instead of directly regressing the point coordinates.
This simple change largely improves the stability of learning, which results in a model with a reasonable performance, as shown by $\textnormal{GCN}^{4\times}$.
We further transform graph convolution into residual graph convolution by putting back the residual connections.
Such a modification improves F-score by about $3\%$ on SHREC15 and $1.5\%$ on the train-test dataset, as shown by $\textnormal{ResGCN}^{4\times}$.
By enabling progressive super-resolution, we obtain the proposed generator.
As shown by ResGCN, the F-score increases by nearly $4\%$ on the train-test dataset.
The Deviation also improves a lot on both datasets.
When changing $\textnormal{L}_{cd}$ into the loss proposed in \cite{yu2018pu}, the F-score decreases by about $8\%$ as shown by ResGCN + $\textnormal{L}_{pu}$ and ResGCN, which demonstrates the effectiveness of $\textnormal{L}_{cd}$.
Compared to PU-Net, ResGCN + $\textnormal{L}_{pu}$ improves the F-score by more than $17\%$ because of the proposed residual graph network ResGCN.
By replacing $\textnormal{L}_{pu}$ with the proposed loss, the F-score is further improved by around $9\%$, which achieves the state-of-the-art performance, as shown by AR-GCN.

We also take an experiment to compare different training strategies.
AR-GCN w/o FT is trained with the proposed loss $L(x, y)$ for $120$ epochs, while AR-GCN is trained with $\textnormal{L}_{cd}$ for 80 epochs and then finetuned with $L(x, y)$ for another 40 epochs.
As shown in the table, AR-GCN outperforms AR-GCN w/o FT consistently, which shows the superiority of the 2-step training strategy.

\begin{table}[!t]
\small
\begin{center}
\begin{tabular}{l|l|c|c|c}
\hline
\multirow{2}{*}{Dataset} & \multirow{2}{*}{Method} & \multicolumn{1}{c}{F-score} & \multicolumn{1}{|c}{NUC} & \multicolumn{1}{|c}{Deviation}\\
\cline{3-5}
& & $\tau=0.01$ & $1.0\%$ & mean \\
\hline
\multirow{4}{*}{Train-Test} & PU-Net & 43.2\% & 0.131 & 0.78 \\
& Ours & 70.3\% & 0.128 & 0.26 \\
\cline{2-5}
& Ours+noisy & 55.4\% & 0.128 & 0.59 \\
\cline{2-5}
& Ours+uneven & 46.0\% & 0.202 & 0.56 \\
\hline
\hline
\multirow{4}{*}{SHREC15} & PU-Net & 56.4\% & 0.180 & 0.90 \\
& Ours & 93.1\% & 0.130 & 0.18 \\
\cline{2-5}
& Ours+noisy & 78.8\% & 0.130 & 0.55 \\
\cline{2-5}
& Ours+uneven & 75.5\% & 0.201 & 0.37 \\
\hline
\end{tabular}
\end{center}
	\vspace{-1.5em}
	\caption{\textbf{Experiments on noisy data and uneven data.}}
	\vspace{-1em}
	\label{table:noisy_uneven}
\end{table}

\subsection{Robustness of AR-GCN}
To show the robustness of our method, we take experiments on noisy point clouds and non-uniformly sampled point clouds separably.
The quantitative results on the Train-Test dataset and SHREC15 are shown in Table~\ref{table:noisy_uneven}.
Ours+noisy represents applying AR-GCN on noisy point clouds (Gaussian noise $z$, where $z\sim \mathcal{N}(0, 0.01)$).
Results show that our method with noisy point clouds outperforms PU-Net with clean ones.
Ours+uneven means employing our method on non-uniformly sampled point clouds.
Our method with uneven point clouds achieves similar performance to PU-Net with uniform ones on all the metrics except NUC.
We attribute this to the non-uniform distribution of the input point clouds.
Our method can only make it more uniform to a certain degree.
We plan to solve this problem in the future work.

\begin{figure}
\captionsetup[subfigure]{labelformat=empty}
\begin{center}
    \subcaptionbox{}{\includegraphics[width=0.25\linewidth]{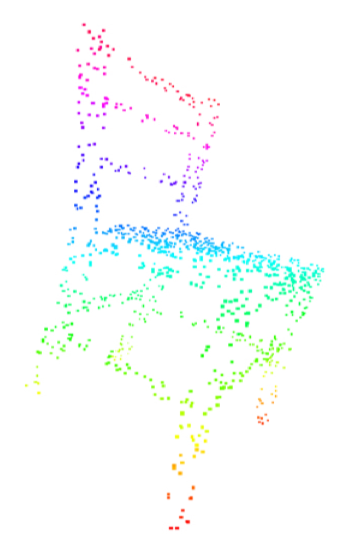}}
    \hfill
	\subcaptionbox{}{\includegraphics[width=0.25\linewidth]{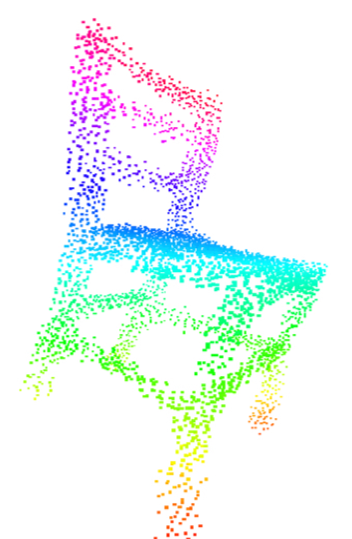}}
	\hfill
	\subcaptionbox{}{\includegraphics[width=0.25\linewidth]{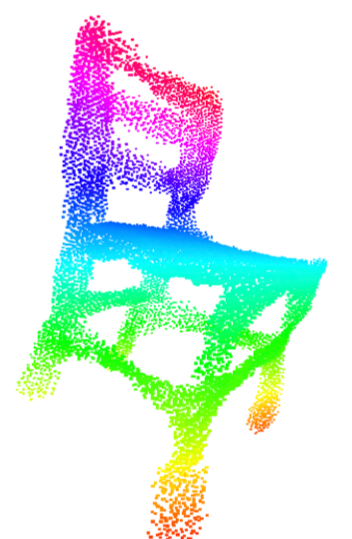}}\\
	\subcaptionbox{(a) Input}{\includegraphics[width=0.25\linewidth]{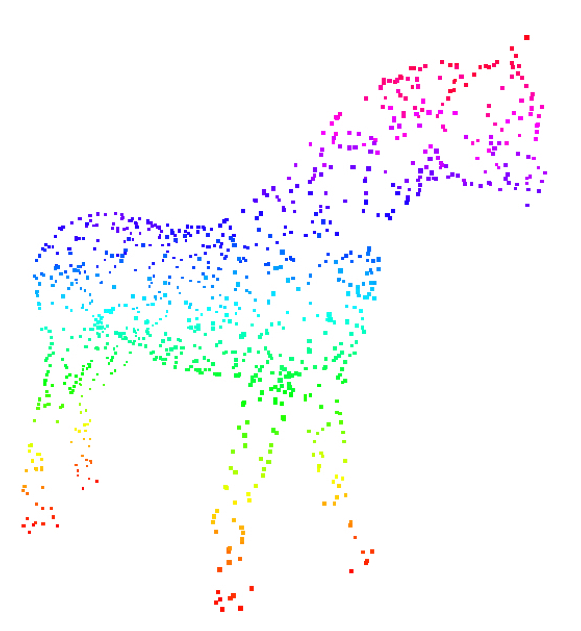}}
	\hfill
	\subcaptionbox{(b) 1st iteration}{\includegraphics[width=0.25\linewidth]{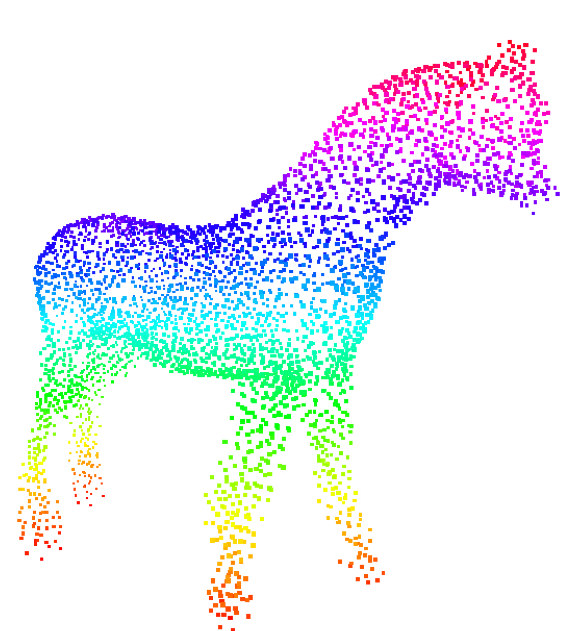}}
	\hfill
	\subcaptionbox{(c) 2nd iteration}{\includegraphics[width=0.25\linewidth]{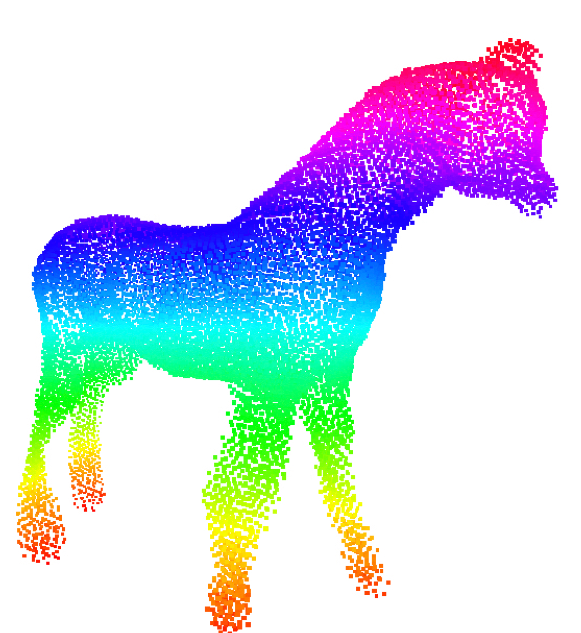}}
\end{center}
	\vspace{-1.5em}
	\caption{\textbf{Iterative Super Resolution.} (a) is the input point cloud with $1,024$ points. (b) and (c) are the generated HR point clouds after the 1st and 2nd iterations. At each iteration, the output from the previous iteration is upsampled by 4 times with our method. Best viewed in color.}
	\vspace{-1em}
	\label{fig:iterative}
\end{figure}

\subsection{Applications}
\paragraph{Iterative Super Resolution}
Our method is trained to upscale a point cloud by a fixed ratio, which is $4\times$ in our setting.
To demonstrate the ability of upsampling a point cloud by more than $4\times$, we conduct an experiment that takes the output of the previous iteration as input and upsamples it by $4\times$ again with AR-GCN.
The initial point cloud has $1,024$ points, which are upsampled by $16\times$ after 2 iterations.
As shown in Figure~\ref{fig:iterative}, our method not only handles a relatively sparse point cloud but also upsamples it by more than $4\times$ iteratively.
Although the resulting point cloud is not as good as that in Figure~\ref{fig:visual}, it recovers many details from the point cloud with only $1,024$ points, which is promising.

\begin{figure}
\begin{center}
    \subcaptionbox{Input}{\includegraphics[width=0.4\linewidth]{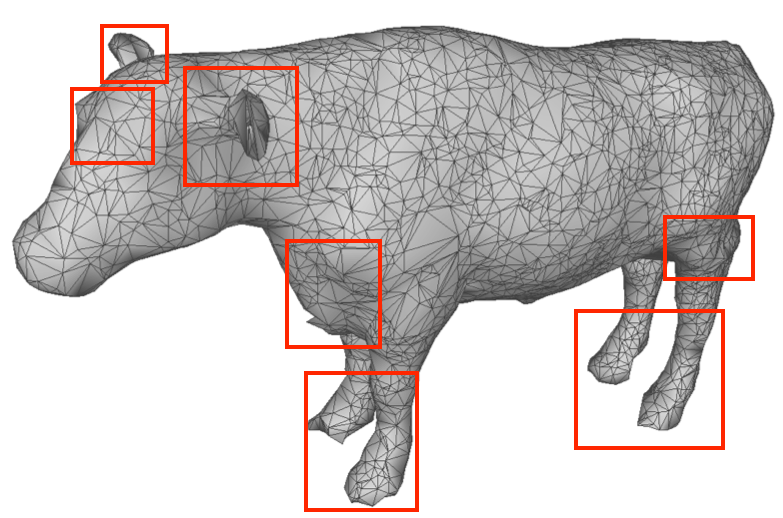}}
    \hfill
	\subcaptionbox{GT}{\includegraphics[width=0.4\linewidth]{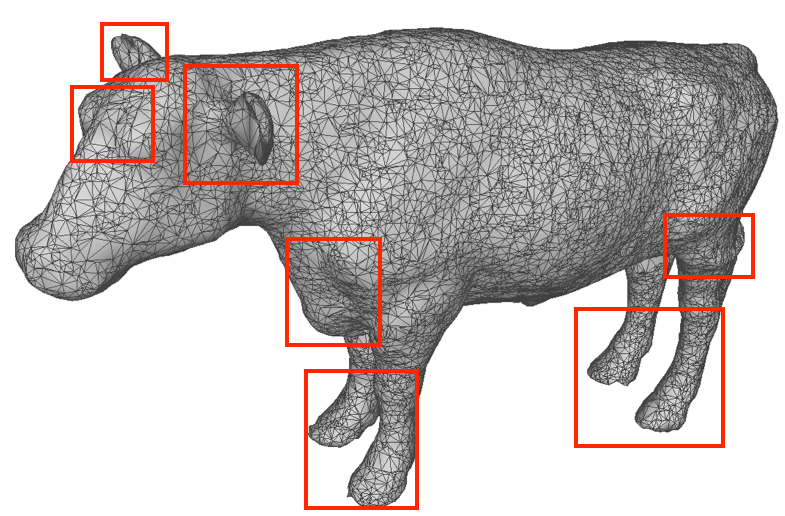}}\\
	\subcaptionbox{PU-Net~\cite{yu2018pu}}{\includegraphics[width=0.4\linewidth]{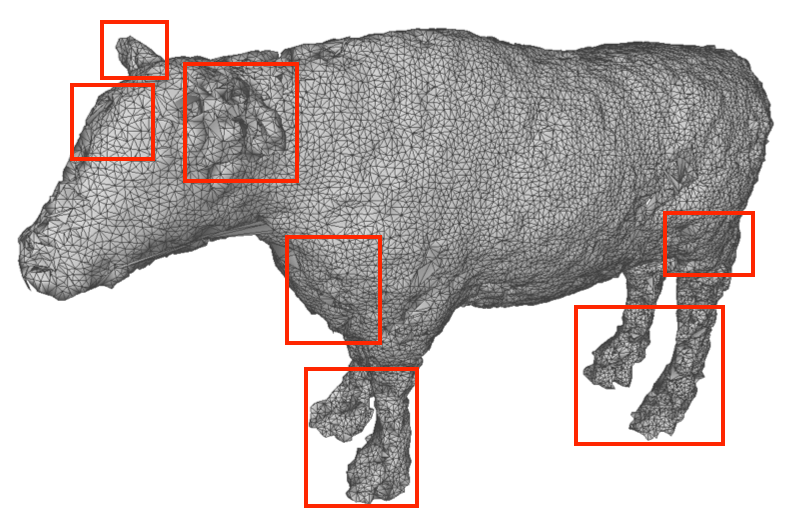}}
	\hfill
	\subcaptionbox{Ours}{\includegraphics[width=0.4\linewidth]{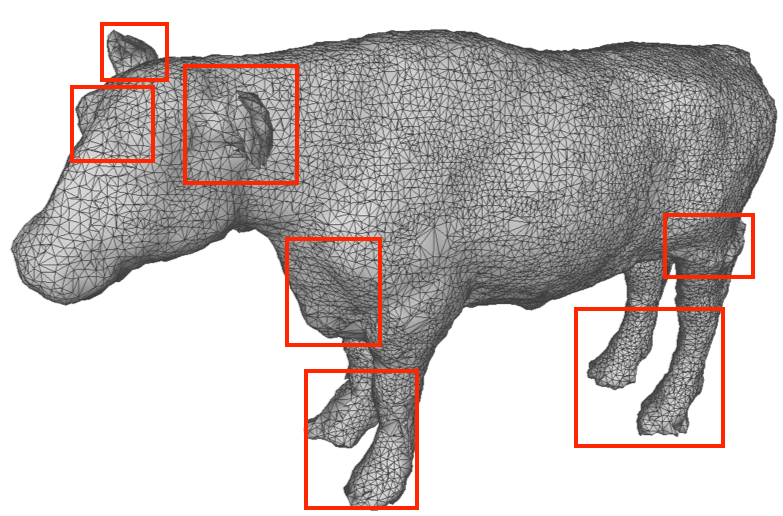}}
\end{center}
	\vspace{-1.5em}
	\caption{\textbf{Mesh Reconstruction from Point Cloud.} The differences inside the red boxes are most obvious. Best viewed in color.}
	\vspace{-1em}
	\label{fig:mesh}
\end{figure}

\vspace{-1em}
\paragraph{3D Reconstruction}
In 3D reconstruction, the quality and density of the point cloud have a huge impact on the quality of the reconstructed mesh.
However, due to the limitations of scanning devices, the point cloud is usually sparse and noisy.
Thus, point cloud super-resolution is the key to improve the quality of 3D reconstruction.
We employ our method and PU-Net to generate the HR point cloud, which is then fed into Ball pivoting algorithm~\cite{bernardini1999ball} for mesh reconstruction.
As shown in Figure~\ref{fig:mesh}, the mesh reconstructed from our method contains richer details compared to that from the LR input.
Besides, ours is smoother in the flat area and sharper at the edges, while the mesh from PU-Net is noisy with many unpleasant artifacts, especially in the red boxes.

\begin{figure}
\begin{center}
    \includegraphics[height=0.45\linewidth]{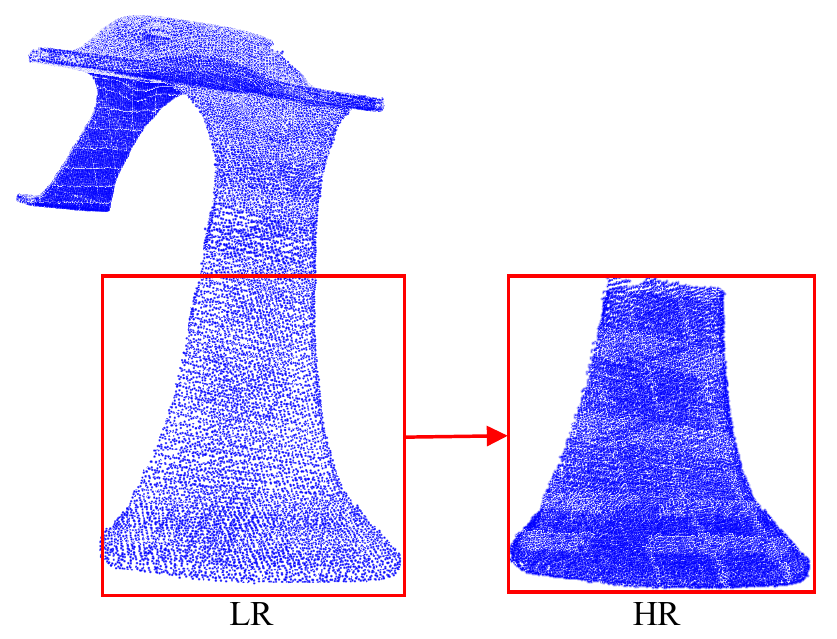}
    \hfill
    \includegraphics[height=0.45\linewidth]{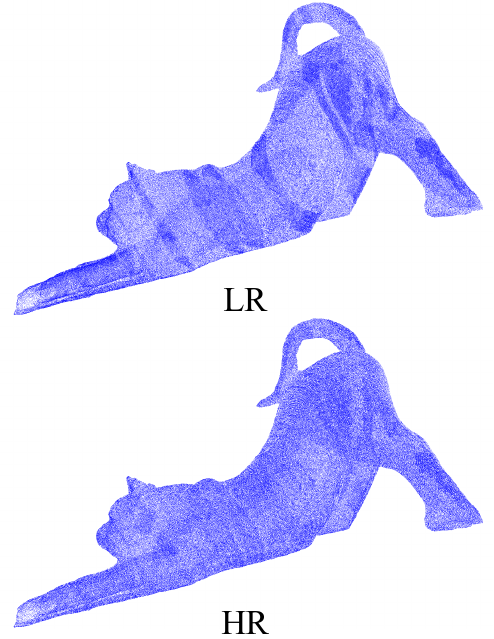}
\end{center}
	\vspace{-1.5em}
	\caption{Employ our method on real-scanned and un-even point cloud. Best viewed in color.}
	\vspace{-1em}
	\label{fig:real}
\end{figure}

\begin{table}[!t]
\begin{center}
\begin{tabular}{l|c|c|c}
\hline
\#Points & 1024 & 256 & 1024 (from 256 points) \\
\hline
Accuracy (\%) & 91.05 & 46.96 & 79.34 \\
\hline
\end{tabular}
\end{center}
	\vspace{-1.5em}
	\caption{Classification accuracy on the test set of ModelNet40 with PointNet++. 1024 (from 256 points) is obtained by upsampling the 256 points 4 times with our method.}
	\vspace{-1em}
\label{table:classification}
\end{table}

\vspace{-1em}
\paragraph{LR Point Cloud Classification}
For 3D understanding, the classification accuracy of the LR point cloud is usually lower than that of the HR point cloud.
To improve the performance of LR point cloud classification, one possible way is transforming it to HR point cloud with the point cloud super-resolution method.
To show the effectiveness of point cloud super-resolution, we employ PointNet++~\cite{qi2017pointnet++} on ModelNet40 dataset~\cite{wu20153d} for point cloud classification.
As shown in Table~\ref{table:classification}, PointNet++ achieves $91.05\%$ in classification accuracy with $1,024$ points as input.
When we randomly sample 256 points from the $1,024$ points and send them to PointNet++, the performance drops from $91.05\%$ to $46.96\%$.
We then upsample the 256 points by $4\times$ with our method and send the $1,024$ points to PointNet++, resulting in $79.34\%$ in accuracy, which outperforms $46.96\%$ by a large margin.
This experiment shows that point cloud super-resolution is important for understanding LR point cloud.

\vspace{-1em}
\paragraph{Real-Scanned Point Cloud}
We also conduct an experiment on real-scanned and un-even point clouds.
As shown in Figure~\ref{fig:real}, our method generates a denser point cloud with more uniformly distributed points, while maintains the underlying structure such as the striped texture.

\section{Conclusion}
We proposed a graph convolution network AR-GCN for point cloud super-resolution, which is composed of a residual generator, a graph discriminator, and a graph adversarial loss.
With comprehensive experiments, we demonstrated that residual connections and residual prediction are effective for stable training and better performance.
With the proposed graph adversarial loss, our method generates more realistic HR point cloud compared to the manually designed loss function.
The experiment on train-test dataset showed that our method outperforms other methods.
The experiment on the unseen dataset SHREC15 further demonstrated the better performance and generalization ability of our method.
Notably, our method is not designed for completion, thus it can not fill large holes or missing parts.
We'd like to solve the limitations in the future work.

{\small
\bibliographystyle{ieee}
\bibliography{egbib}
}

\end{document}